\documentclass[aps,prb,10pt,onecolumn,superscriptaddress,nobibnotes]{revtex4-2}
\pdfoutput=1

\usepackage{lmodern}
\usepackage[T1]{fontenc}
\usepackage[utf8]{inputenc}
\usepackage{graphicx}
\usepackage{amsmath}
\usepackage[unicode=true,
            colorlinks=true,
            linkcolor=blue,
            citecolor=blue,
            urlcolor=blue]{hyperref} 
\usepackage{siunitx}
\usepackage{multirow}
\usepackage{upgreek}

\usepackage{colortbl}
\usepackage{footnote}
\usepackage{amssymb}			
\usepackage{hyperref}		
\usepackage[flushleft]{threeparttable}

\usepackage{float}

\usepackage{subfig}

\usepackage{footnote}
\usepackage{booktabs,caption}
\usepackage[flushleft]{threeparttable}
\usepackage[bottom]{footmisc}				

\usepackage[x11names]{xcolor}

\newcommand{\domark}{%
  \vbox to 0pt{
    \kern-\dp\strutbox
    \smash{\llap{\color{red!90!black}\#\kern0.5em}}
    \vss
  }%
}

\immediate\write18{git rev-parse --short HEAD > \jobname.gitinfo }
\newcommand{\gitrev}{\InputIfFileExists{\jobname.gitinfo}{}{ref not available}}

\begin{document}

\title{High-Speed Graphene-based Sub-Terahertz Receivers enabling Wireless Communications for 6G and Beyond}

\author{Karuppasamy Pandian Soundarapandian} \affiliation{ICFO - Institut de Ci\`{e}ncies Fot\`{o}niques, The Barcelona Institute of Science and Technology, 08860, Castelldefels (Barcelona), Spain} \affiliation{These authors contributed equally}
\author{Sebasti\'{a}n Castilla} \email{sebastian.castilla@icfo.eu} \affiliation{ICFO - Institut de Ci\`{e}ncies Fot\`{o}niques, The Barcelona Institute of Science and Technology, 08860, Castelldefels (Barcelona), Spain} \affiliation{These authors contributed equally}
\author{Stefan M. Koepfli} \affiliation{ETH Zurich, Institute of Electromagnetic Fields (IEF), 8092 Zurich, Switzerland}
\author{Simone Marconi} \affiliation{ICFO - Institut de Ci\`{e}ncies Fot\`{o}niques, The Barcelona Institute of Science and Technology, 08860, Castelldefels (Barcelona), Spain}
\author{Laurenz Kulmer} \affiliation{ETH Zurich, Institute of Electromagnetic Fields (IEF), 8092 Zurich, Switzerland}
\author{Ioannis Vangelidis} \affiliation{Department of Materials Science and Engineering, University of Ioannina, 45110 Ioannina, Greece}
\author{Ronny de la Bastida} \affiliation{Catalan Institute of Nanoscience and Nanotechnology (ICN2), BIST and CSIC, Campus UAB, 08193 Bellaterra (Barcelona), Spain}
\author{Enzo Rongione} \affiliation{Catalan Institute of Nanoscience and Nanotechnology (ICN2), BIST and CSIC, Campus UAB, 08193 Bellaterra (Barcelona), Spain}
\author{Sefaattin Tongay} \affiliation{Materials Science Division, Lawrence Berkeley National Laboratory, Berkeley, 94720, California, USA} \affiliation{Department of Materials Science and Engineering, University of California, Berkeley, 94720-1760, California, USA} \affiliation{School for Engineering of Matter, Transport and Energy, Arizona State University, Tempe, Arizona 85287, USA}
\author{Kenji Watanabe} \affiliation{Research Center for Electronic and Optical Materials, National Institute for Materials Science, 1-1 Namiki, Tsukuba 305-0044, Japan}
\author{Takashi Taniguchi} \affiliation{Research Center for Materials Nanoarchitectonics, National Institute for Materials Science, 1-1 Namiki, Tsukuba 305-0044, Japan}
\author{Elefterios Lidorikis} \affiliation{Department of Materials Science and Engineering, University of Ioannina, 45110 Ioannina, Greece} \affiliation{University Research Center of Ioannina (URCI), Institute of Materials Science and Computing, 45110 Ioannina, Greece}
\author{Klaas-Jan Tielrooij} \affiliation{Catalan Institute of Nanoscience and Nanotechnology (ICN2), BIST and CSIC, Campus UAB, 08193 Bellaterra (Barcelona), Spain} \affiliation{Department of Applied Physics, TU Eindhoven, Den Dolech 2, Eindhoven 5612 AZ, Netherlands}
\author{Juerg Leuthold} \affiliation{ETH Zurich, Institute of Electromagnetic Fields (IEF), 8092 Zurich, Switzerland}
\author{Frank H.L. Koppens} \email{frank.koppens@icfo.eu} \affiliation{ICFO - Institut de Ci\`{e}ncies Fot\`{o}niques, The Barcelona Institute of Science and Technology, 08860, Castelldefels (Barcelona), Spain} \affiliation{ICREA - Instituci\'o Catalana de Recerca i Estudis Avan\c{c}ats, 08010 Barcelona, Spain}

\maketitle

\section*{\textsf{Abstract}}
{
\bf
In recent years, the telecommunications field has experienced an unparalleled proliferation of wireless data traffic. Innovative solutions are imperative to circumvent the inherent limitations of the current technology, in particular in terms of capacity. Carrier frequencies in the sub-terahertz (sub-THz) range ($\sim$0.2-0.3 THz) can deliver increased capacity and low attenuation for short-range wireless applications. Here, we demonstrate a direct, passive and compact sub-THz receiver based on graphene, which outperforms state-of-the-art sub-THz receivers. These graphene-based receivers offer a cost-effective, CMOS-compatible, small-footprint solution that can fulfill the size, weight, and power consumption (SWaP) requirements of 6G technologies. We exploit a sub-THz cavity, comprising an antenna and a back mirror, placed in the vicinity of the graphene channel to overcome the low inherent absorption in graphene and the mismatch between the areas of the photoactive region and the incident radiation, which becomes extreme in the sub-THz range. The graphene receivers achieve a multigigabit per second data rate with a maximum distance of $\sim$3 m from the transmitter, a setup-limited 3 dB bandwidth of 40 GHz, and a high responsivity of 0.16 A/W, enabling applications such as chip-to-chip communication and close proximity device-to-device communication.\\
}

Wireless communications have made remarkable strides with the advent of 5$^{\rm th}$ generation (5G) technology, enabling unprecedented levels of connectivity and opening up new possibilities for innovation and progress in various fields. Following Edholms' law, data traffic has doubled every 18 months over the last three decades. This trend predicts a  data speed requirement of terabit per second (Tbits$^{-1}$) before 2035 \cite{kraemer_2017} to meet the rising demands. This has led to the development of 6G, which targets extremely high data speeds ($\sim$1 Tbits$^{-1}$), ultra-low latency (below a ms)\cite{you_2021}, and advanced wireless connectivity, which are mandatory in the fields of medical robotics, surveillance drones, and digital-physical twins \cite{bassoli_2021, bouchmal_2023}. The transition from 5G to 6G will include the increase of carrier frequency from microwave to certain bands in the terahertz (THz) range, as this offers higher bandwidth and faster data transfer rates \cite{koenig_2013,sengupta_2018, khudchenko_2023, guerboukha_2024, alimi_2024}. Operation at sub-THz frequencies (0.2-0.3 THz) is considered optimal because of the significant atmospheric attenuation above 1 THz. \cite{nagatsuma_2016,itug_2005}.
\\

Receivers play a crucial role in wireless communication approaches for detecting data signals. State-of-the-art sub-THz receivers can loosely be categorized into electrical and optical receivers. Electrical based schemes such as heterodyne detection\cite{heller_2016,yangieee_2014,simsekieee_2014,yang_155_2014} achieve data rates in the 120 Gbits$^{-1}$ range\cite{koenig_2013,yangieee_2014, hamada_2020, wrana_2022, wrana_2023}. They are typically constrained by their relatively non-flat frequency response, bandwidth and non-linearities due to the employed mixers. On the other hand, the optical receivers employ Schottky barrier diodes (SBD) to detect the sub-THz radiation and directly convert it to the baseband. Such devices have been used to demonstrate $\sim$100 Gbits$^{-1}$ data transmission\cite{harter_2020,tohme_2014,jastrow_2008,Yadav_2023, rogalski_2022}. However, they are bulky and offer limited bandwidths. Recently, photonic receivers based on high-speed plasmonic modulators have been demonstrated. Here the device offers a bandwidth up to 500 GHz, and beyond\cite{heni_2016, burla_2019,salamin_2018, horst_2022, blatter_2024, ibili_2024}. The plasmonic modulator encodes the data from the sub-THz carrier onto a telecommunication optical carrier to detect the data signal with a conventional photodetector. High data rates $>$100 Gbits$^{-1}$ over distances $>$1 km have been demonstrated with this approach. See Table I for a summary of key parameter values.
\\
 
While all of the above-mentioned approaches are capable of operation in the sub-THz regime and achieve high data rates, they still involve high complexity due to harmonic mixers, local oscillators, amplifiers, additional photodetectors, etc. This leads to challenging integration in terms of footprint and power consumption. These limitations prevent the scalability and diverse integration of these receivers. Moreover, none of the current techniques offer simultaneously the following desired characteristics\cite{song_2011,testa_2022}: low noise, 50 $\Omega$ output impedance, compatibility with complementary metal-oxide-semiconductor (CMOS) technology \cite{gossens_2017}, low cost, and small footprint.
\\

Here, we demonstrate that graphene-based receivers are promising candidates for 6G wireless technology \cite{Montanaro_2023, Montanaro_2024}, as they simultaneously meet all the above-mentioned requirements. In particular, we show the detection of data streams using sub-THz graphene receivers. Our graphene-based receivers are based on a direct detection scheme, where incident data with a sub-THz carrier frequency is directly converted into electronic signals, without the need for a local oscillator or an upconversion process. These receivers can operate in a passive mode by exploiting the photothermoelectric (PTE) effect in graphene at room temperature and zero-bias operation\cite{cai_2014, tielrooij_2018, castilla_2019, muench_2019, marconi_2021}. This approach ensures zero dark current, direct photovoltage or photocurrent readouts, low flicker noise, and compatibility with CMOS platforms. Consequently, it allows for a simple circuitry design and low power consumption, marking a significant milestone in the field of wireless communications. 
\\

The capabilities of this sub-THz graphene direct receiver are highly promising. Recently, an intrinsic 3 dB bandwidth exceeding 400 GHz was demonstrated for a photothermoelectric (PTE)-based graphene detector operating at telecom frequencies \cite{koepfli_2024}, suggesting the potential for data rates exceeding 500 Gbits$^{-1}$. Furthermore, our simulations predict that responsivities of approximately 1 A/W (120 V/W) and a noise-equivalent power (NEP) of 14 pW/$\sqrt{\rm Hz}$ are attainable. These predictions are supported by experimental results, where we achieved a multigigabit per second data rate over a distance of approximately 3 meters between the transmitter and the graphene receiver. We demonstrated a setup-limited 3 dB bandwidth of 40 GHz, a responsivity of 0.16 A/W (30 V/W), and an NEP of 58 pW/$\sqrt{\rm Hz}$. These results highlight the potential of the receiver for applications such as chip-to-chip communication and close proximity device-to-device communication.
\\

\section*{\textsf{R\lowercase{esults}}}
\begin{figure*} [t]
	\includegraphics[width=\textwidth]
	{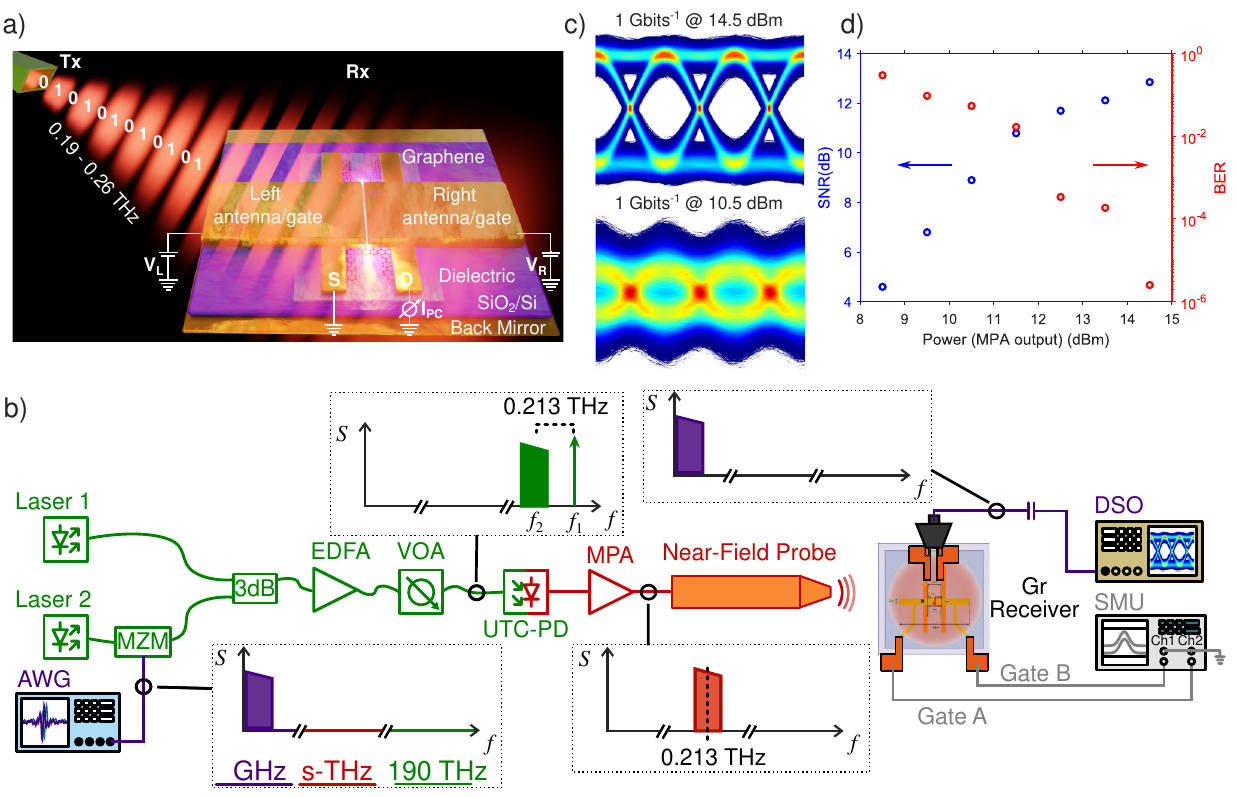}
	\caption{ 
		\footnotesize \textbf{Sub-THz digital data-stream detection with graphene-based receiver}. \textbf{a)} Schematic representation of a typical configuration of receivers with incident sub-THz radiation including the device circuitry. The illustration contains the back mirror, Si/SiO$_2$ substrate, graphene encapsulated with dielectrics (e.g., hBN, alumina), and antenna/split-gate.
		\textbf{b)} Experimental setup used to detect the data stream. In this case, we employ a near-field probe to compensate the sub-THz beam divergence. The panels in dashed lines show the corresponding frequency of the utilized signal in each component, where purple, orange, and green correspond to the electrical (radiofrequency, RF), sub-THz, and telecom frequencies, respectively.
		\textbf{c)} Eye diagrams of NRZ OOK data streams at 1 Gbits$^{-1}$ at different output powers of the transmitter (estimated at the output of the multiport power amplifier, MPA) of 14.5 dBm (top panel) and 10.5 dBm (bottom panel). The carrier frequency is 0.213 THz.
		\textbf{d)} The signal-to-noise ratio (SNR) and bit error rate (BER) extracted from the collected eye diagrams at 1 Gbits$^{-1}$ at different output power of the transmitter.
	}
	\label{fig_1}
\end{figure*}


Fig. \ref{fig_1}a shows the graphene-based receiver, which consists of an hBN encapsulated graphene. The optimized geometry comprises a short and wide graphene channel (see Methods for details) that exhibits reduced resistance that nearly matches the 50 Ohm impedance of the measurement electronics. The receivers operate at zero-bias by exploiting the PTE effect, where the sub-THz radiation absorbed by graphene generates a photoresponse driven by a temperature gradient, provided there is doping asymmetry in the channel\cite{mics_2015,castilla_2019,castilla_2020, vaidotas_2020, viti_2021, vangelidis_2022, castilla_2024}. In our implementation, the doping asymmetry across the graphene channel is induced using a split-gate structure. Independent voltages are applied to the left and right gates to create a graphene pn-junction, thereby maximizing the Seebeck coefficient difference between the n-doped and p-doped regions and, consequently, enhancing the photoresponse\cite{castilla_2019, castilla_2020}, as detailed in Supplementary Note 3.
\\

This split-gate simultaneously serves as a half-wavelength dipole antenna, with a length of 600 \textmu m, to attain a resonant frequency of $\sim$0.230 THz. The sub-THz radiation is highly concentrated at the $\sim$400 nm gap of the antenna, which overlaps with the photoactive area (i.e., the graphene pn-junction)\cite{castilla_2019}. In conjunction with this, we employ a back mirror below the highly resistive silicon substrate, which is transparent in the sub-THz range and has a thickness of $\sim$300 \textmu m, as shown in Fig. \ref{fig_1}a. By following this approach, we achieve multiple reflections between the back mirror and antenna, where the graphene monolayer is located to significantly boost the absorption. Thus, the integrated antenna and a back mirror enable to overcome the low graphene absorption and large mismatch between the small optically active area ($\sim$\textmu m$^2$) with respect to the considerable large spot size that becomes extreme at sub-THz wavelengths ($\sim$mm$^2$)\cite{castilla_2019,hafez_terahertz_2020,karpowicz_non-destructive_2005}.
\\

We have fabricated 7 devices, where 2 are based on exfoliated graphene and 5 on graphene grown by chemical vapor deposition (CVD); see more details in Supplementary Notes 1-3. We also prepared an evaluation kit with a packaged THz receiver containing one of the devices based on exfoliated graphene. We will first discuss the results obtained with the device based on exfoliated graphene labeled as Device 1 (D1).
\\


We evaluate the data stream detection of Device 1, which consists of hBN encapsulated exfoliated graphene, as depicted in Fig. \ref{fig_1}a. Fig. \ref{fig_1}b shows the experimental configuration that we used for sub-THz data stream detection (more details in Supplementary Note 4). We use two lasers, denoted as lasers 1 and 2 in Fig. \ref{fig_1}b, which operate at distinct frequencies (\(f_1\) and \(f_2\), respectively) in the near-infrared range ($\sim$1550 nm) and are combined using a 3 dB splitter. We control the detuning of these two lasers (\(f_1 - f_2\) $\approx$ 0.213 THz) to generate a beating tone, which in the high-speed uni-travelling carrier photodiode (UTC-PD) is detected\cite{horst_2022} and is transmitted to a near-field probe. The fabricated sub-THz receiver is placed at a distance of a few centimeters from the near-field probe output. We employ a near-field probe to get more radiation onto the device, because the sub-THz beam is strongly divergent. The output signal from the graphene receiver was amplified by a voltage amplifier with a variable gain ranging from 30 dB to 70 dB and a 3 dB bandwidth of 1.2 GHz and a real-time digital sampling oscilloscope (DSO) was employed to record the sampled waveforms.
\\ 

To encode data onto the sub-THz carrier, we use a modulator and an arbitrary waveform generator on the optical carrier generated by laser 2, see Figure \ref{fig_1}b. We encode a random bit sequence with non-return-to-zero on-off keying (NRZ OOK) modulation format at 1, 1.5, and 2 Gbits$^{-1}$. The collected eye diagrams are shown in Fig. \ref{fig_1}c and Supplementary Note 5. The completely open eye diagram at 1 Gbits$^{-1}$ shows that we successfully convert sub-THz wireless data-streams into electronic signals. The eye becomes narrower upon increasing the bit rate above 1 Gbits$^{-1}$, and it completely closes for a bit rate of 3 Gbits$^{-1}$, as shown in Supplementary Note 5. This is setup-limited and arises from the 1.2 GHz bandwidth of the voltage amplifier connected to the DSO.
\\ 

Subsequently, we measure the eye diagrams for different input powers (measured at the output of the multiport power amplifier, MPA), which is achieved by tuning the variable optical attenuator (VOA), as shown in Fig. \ref{fig_1}b. We observe that the amplitude of the eye decreases with power, as shown in Fig. \ref{fig_1}c. The extracted bit error rate (BER) from the detected data streams at different powers is shown in Fig. \ref{fig_1}d. The BER reaches a minimum value of $\sim$2.5$\times$10$^{-6}$. This minimum BER value was setup limited by the memory of used instruments. For the signal-to-noise ratio (SNR) as a function of power, we achieve a maximum value of $\sim$13 dB and a minimum of 4 dB for the lowest power. These results were obtained while maintaining a constant voltage amplifier gain of 40 dB. In Supplementary Note 5, we also evaluate the BER and SNR as a function of the amplifier gain by varying it from 30 dB to 70 dB. Knowing these parameters, such as the optimum amplification gain, enables the design of evaluation kits or future prototypes with an integrated amplifier featuring optimal gain for our sub-THz receivers.  
\\

\begin{figure*} [t]
	\includegraphics[width=\textwidth]
	{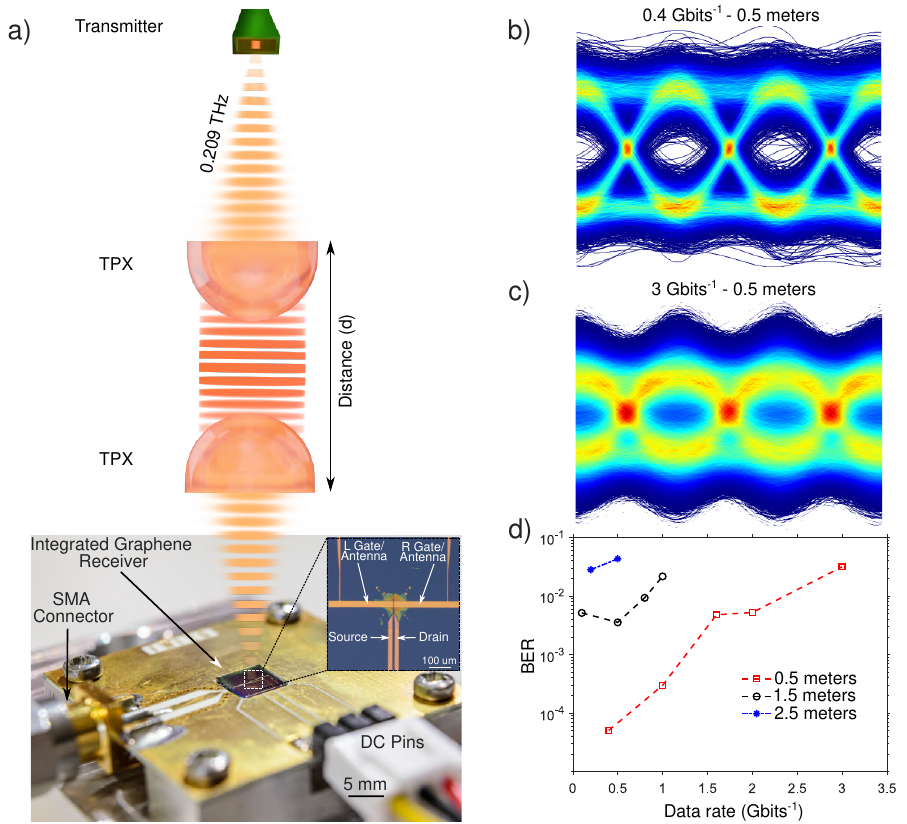}
	\caption{ 
		\footnotesize \textbf{Evaluation kit and digital data-stream detection as a function of the sub-THz wireless link distance}. 
		\textbf{a)} Schematic of the configuration containing lenses with a variable distance between the transmitter and the graphene receiver. On the bottom part, the optical image of the side view of the fabricated PCB-integrated receiver on a metal support. The inset shows the optical image of the graphene receiver.
		\textbf{b)} Collected eye diagrams at data rates of 0.4 and \textbf{c)} 3 Gbits$^{-1}$ with a distance of 0.5 m between the lenses and 0.64 m between the transmitter and receiver. The carrier frequency is 0.209 THz.
		\textbf{d)} BER in log scale as a function of data rate detection for different distances of the sub-THz wireless link.
	}
	\label{fig_2}
\end{figure*}



\begin{figure*} [t]
	\includegraphics[width=\textwidth]
	{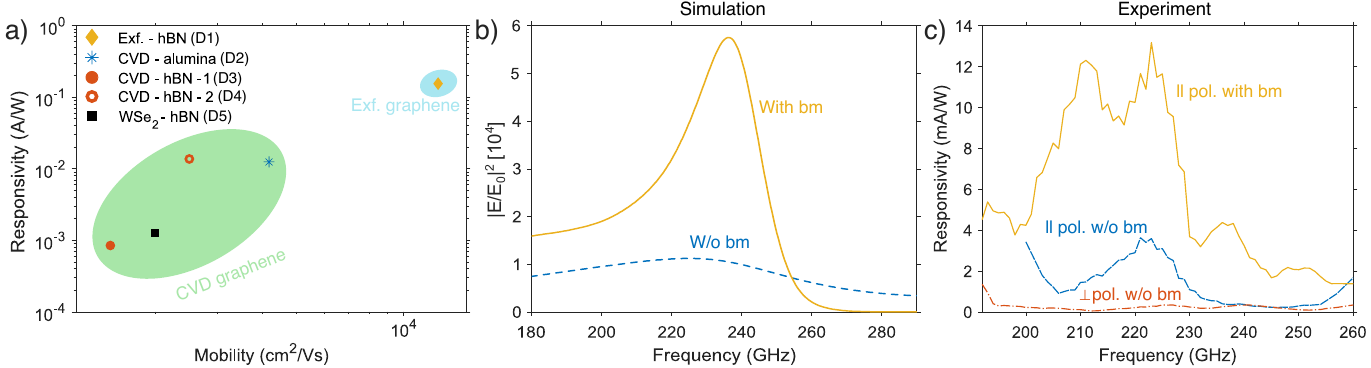}
	\caption{ 
		\footnotesize \textbf{Spectral responsivity of the sub-THz receivers}. 
		\textbf{a)} Responsivity of the investigated devices as a function of the charge carrier mobility of graphene. The highest responsivity is observed for the hBN-encapsulated exfoliated graphene receiver, which exhibits the highest mobility.
		\textbf{b)} Optical simulations of the field intensity at the graphene channel with and without the back mirror (bm) within the measured spectral range.
		\textbf{c)} Responsivity spectra of Device 2 (D2), which is based on CVD graphene with alumina as the dielectric. The highest responsivity is achieved in the presence of the back mirror and with the incident sub-THz radiation polarization parallel to the antenna main axis.
	}
	\label{fig_3}
\end{figure*}



Following the assessment of these initial data stream detection measurements with sub-THz carrier frequencies, we developed a graphene-based wireless receiver evaluation kit for the sub-THz range and use it to demonstrate a wireless link over different distances. The evaluation kit receiver, consists of a printed circuit board (PCB) with an integrated hBN-encapsulated exfoliated graphene receiver, similar to that of Device 1. In this configuration, the receiver is positioned at the center of the PCB (as indicated in Fig. \ref{fig_2}a and Supplementary Note 6), with DC pin heads for the split gates, and the ground plane is situated on the right side of the PCB. From the middle to the bottom of the PCB, the microstrip transmission line, designed to have a characteristic impedance of 50 $\Omega$, carries the photogenerated signal that terminates with an SMA connector. A metal block serves as both a support and back mirror (see Fig. \ref{fig_2}a).
\\

Using a similar NRZ OOK amplitude modulation scheme, with a carrier frequency of 209 GHz, we study data transfer using the evaluation kit. We implement three main differences compared to the previous measurements: i) the use of two polymethylpentene (TPX) lenses for collimating and focusing the sub-THz radiation from the horn antenna with a gain of 31 dB, replacing the near-field probe; ii) with the PCB configuration, we avoid the use of RF probes and potential light reflections produced by them; and iii) we vary the distance (\(d\)) between the two TPX lenses from 0.5 m to 2.5 m as shown in Fig. \ref{fig_2}a. The corresponding free space losses are 42, 50 and 54 dB for 0.64, 1.64, and 2.64 m respectively, which correspond to the distances between the transmitter and the graphene receiver with a dipole antenna gain of 2 dB. This setup allows the examination of data transfer as a function of distance.
\\

In Figs. \ref{fig_2}b-c, we show the collected eye diagrams for the data bit rates at 0.4 and 3 Gbits$^{-1}$ for a wireless link with a distance of 0.5 m between the lenses, corresponding to a distance of 0.64 m between transmitter and receiver. Fig. \ref{fig_2}d summarizes the BER measured with respect to data rates, clearly exhibiting an increase of BER and a drop in the maximum measurable data rate upon increasing the distance ($d$) between the transmitter and graphene receiver. These results indicate a reduction in the sub-THz incident power owing to free-space losses and beam divergence; therefore, a higher BER is observed even for small bit rates at the largest measured distance. Consequently, it is more difficult to measure higher bit rates at larger distances. A potential solution to overcome this limitation is to further enhance the responsivity of graphene receivers and the integration of sub-THz cavities. In the following sections, we explain how to increase the responsivity of devices and if this leads to bandwidth limitations. We note that we measured a maximum data rate of 3 Gbits$^{-1}$ at $d =$ 0.5 m, and up to 0.5 Gbits$^{-1}$ at a distance of 2.5 m. In the case of BER, we achieved minimum values of $\sim$2$\times$10$^{-5}$ and $\le$3$\times$10$^{-2}$ at a distance of 0.5 m and 2.5 m respectively.
\\


Now that we have demonstrated a sub-THz wireless link using a graphene-based receiver evaluation kit, we discuss how these devices can be improved and made scalable. We have fabricated 4 additional CVD graphene-based devices by following the same design as Device 1. We use various dielectrics or bottom encapsulants such as alumina, hBN, or transition metal dichalcogenide (TMD) to evaluate their photodetection performance. All the characteristics of the devices are described in the Supplementary Note 2 and Methods.
\\

Fig. \ref{fig_3}a shows a summary of the measured responsivity (in A/W units) of the fabricated sub-THz receivers as a function of the graphene mobility. We note that the higher mobility of graphene enables higher responsivity and lower noise-equivalent power (NEP) because of the improved electronic properties and lower residual doping (n$^*$)\cite{ayush_2022, zihlmann_2019}, which leads to a larger Seebeck coefficient\cite{muench_2019, castilla_2020, asgari_2021}. As a result, we enhance the PTE effect and overall photoresponse in graphene\cite{castilla_2019,muench_2019}. Hence, the hBN-encapsulated exfoliated graphene receiver exhibits the highest external responsivity of 0.16 A/W (30 V/W) with a low NEP of 58 pW/\(\sqrt{\text{Hz}}\), limited only by Johnson noise due to the zero-bias operation\cite{castilla_2019, viti_2021}. These results are consistent with the calculated responsivity values (see Supplementary Notes 7-8). On the other hand, the receiver containing scalable materials (e.g., Device 2, D2) shows a responsivity of 13 mA/W (7.1 V/W) with a NEP of 422 pW/\(\sqrt{\text{Hz}}\). We stress that the detection of data streams by CVD graphene-based receivers is not yet achievable owing to their lower responsivity. These results demonstrate that the high quality of graphene significantly enhances its photoresponse in the sub-THz range\cite{castilla_2019}.
\\

We now discuss how we ensure an enhanced absorption of the incident sub-THz radiation. We simulate our entire device using finite-difference time-domain (FDTD), as described in Methods and Supplementary Note 7. The resonant peak is spectrally located at approximately $\sim$235 GHz, corresponding to a wavelength of $\sim$1.2 mm. This resonance is expected based on the half-wavelength dipole antenna with a length of 600 \textmu m. From simulations with and without the back mirror, we find that the field intensity increases by a factor of $\sim$6 when the back mirror is integrated into the device, as shown in Fig. \ref{fig_3}b.
\\

To evaluate the enhancement provided by the sub-THz cavity, we then measure the frequency-dependent photoresponse of a CVD graphene-based receiver with alumina as the dielectric (labeled Device 2, D2), as depicted in Fig. \ref{fig_3}c. We notice that when having the incident radiation perpendicularly polarized to the main axis of the antenna, we obtain a very low responsivity below 1 mA/W at $\sim$220 GHz. However, when exciting the dipole antenna with the incident polarization parallel to its main axis, we achieve an increase by a factor of 15. The responsivity is further enhanced by placing a metallic back mirror below the Si substrate, thus attaining a boost by factors of 4 and 60 compared to the cases without a back mirror and when the antenna is either excited or not, respectively. The spectral position of the resonant peak and the enhancement provided by the back mirror are in excellent agreement with the simulation results.
\\

\begin{figure*} [t]
	\includegraphics[width=\textwidth]
	{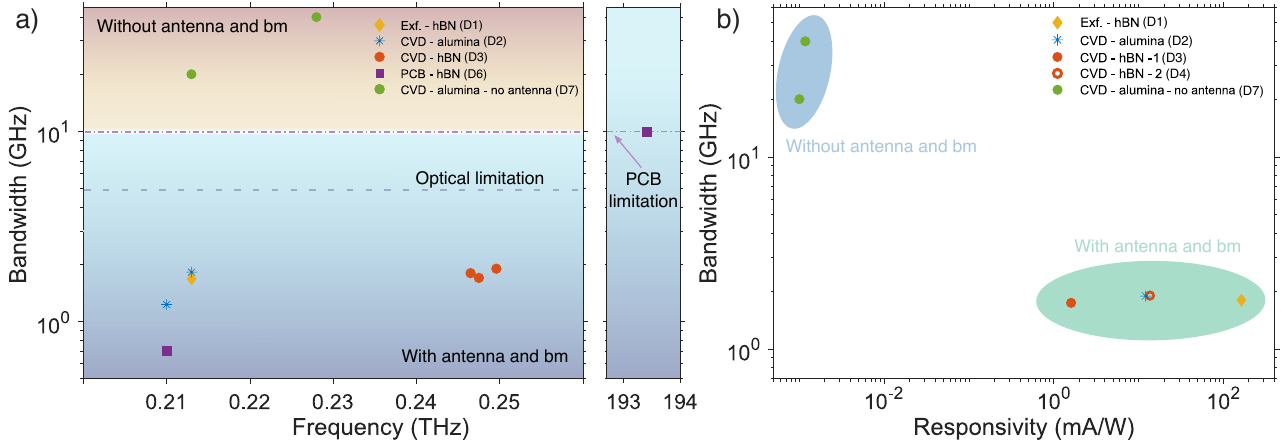}
	\caption{ 
		\footnotesize \textbf{Bandwidth-responsivity trade-off of receivers at different carrier frequencies}.
		\textbf{a)} Bandwidth measurements of all investigated devices as a function of the carrier frequency, including the sub-THz range and the telecom frequency ($\sim$193.5 THz). Devices containing (not containing) a sub-THz antenna and back mirror are highlighted in a blue (red) background. The blue dashed lines indicate the optical bandwidth limitation range, whereas the purple dashed line indicates the bandwidth limitation of the PCB microstrip.
		\textbf{b)} Bandwidth measurements as a function of responsivity in the sub-THz range of the graphene receivers. Devices that contain or do not contain a sub-THz antenna and back mirror are highlighted.
	}
	\label{fig_4}
\end{figure*}


Finally, we study the intrinsic speed limitations that affect these graphene-based sub-THz receivers, including the setup electronics, device resistor-capacitor (RC) time constant, and sub-THz antenna-cavity response. By considering the resistance (R) and capacitance (C) of the devices and the equivalent circuit\cite{marconi_2021, Xia_2009}, we can determine the expected RC-limited 3 dB bandwidth using the following equation: $\approx$1/(2$\pi$RC)\cite{marconi_2021, Xia_2009, castilla_2020}, where the obtained values are $>$50 GHz (see Supplementary Notes 9-10).
\\

On the other hand, the optical limitation of the bandwidth can be calculated by the equation $\approx$\( \frac{c}{n2LQ} \), where \( c \) is the speed of light, $n$ is the refractive index of the substrate (3.42 for silicon), \( L \) is the length of the antenna, and \( Q \) is the quality factor derived from the resonance width\cite{doukas_2018, yu_2015}. Considering an antenna length of 600 \textmu m and a Q-factor of approximately 15 for the measured spectral photoresponse peak (see Fig. \ref{fig_3}b and Supplementary Notes 7 and 10), which corresponds to a bandwidth of $\sim$5 GHz. We note that a similar optical bandwidth limitation is observed for a graphene receiver coupled to a microring resonator at telecom frequencies.\cite{Schuler_2021}
\\

To evaluate the 3 dB bandwidth of the fabricated receivers as a function of carrier frequency, we use a setup similar to that described in Fig. \ref{fig_1}b (see Supplementary Note 4 for further details of the setup). In Fig. \ref{fig_4}a, the sub-THz receivers show a limited 3 dB bandwidth ranging between $\sim$1 and 2 GHz at frequencies between 0.21 and 0.25 THz. In addition, by varying the gate voltage (carrier density), which affects the graphene's resistance and capacitance,\cite{Xia_2009} we do not observe significant changes in the bandwidth values, as shown in Supplementary Fig. 22. However, when testing the receiver integrated with the PCB (Device 6) at a wavelength of 1550 nm (193.4 THz), it shows a bandwidth of 10 GHz, which is limited by the microstrip line of the PCB, as shown in Supplementary Fig. 23. These results demonstrate that, despite having the same capacitance and arrangement of source-drain electrodes, the sub-THz receiver exhibits a carrier-frequency-dependent bandwidth.
\\

We fabricate and test a 7$^{\rm th}$ device (D7) that consists of a CVD graphene-based sub-THz receiver with a non-resonant split-gate architecture, that is, with a short split-gate length ($\sim$10 \textmu m), a graphene channel of approximately the same length and without a back mirror. In contrast to the other devices, Device 7 exhibits a bandwidth response of up to 40 GHz (limited by the Mach-Zehnder modulator, MZM) at a carrier frequency of 0.23 THz. Therefore, these results reinforce the optical bandwidth limitation imposed by the sub-THz cavity in previous devices and demonstrate that higher bandwidths can be achieved with graphene-based sub-THz receivers.
\\

Next, we evaluate the bandwidth of the devices as a function of the measured responsivity in Fig. \ref{fig_4}b. We observe that the sub-THz receivers with a responsivity higher than 1 mA/W, which contain the antenna and back mirror, show a limited 3 dB bandwidth ranging between 1 and 2 GHz. Although Device 7 exhibits a very low responsivity of $\sim$10 \textmu A/W, its bandwidth reaches setup-limited values of 20-40 GHz. Our findings suggest a crucial trade-off between bandwidth and responsivity in these receivers.
\\

\section*{\textsf{D\lowercase{iscussion}}}

\newcolumntype{s}{>{\columncolor[HTML]{AAACED}} p{2cm}}
\setlength{\arrayrulewidth}{0.1mm}

\begin{savenotes}
	\begin{table*}[ht]
		\begin{threeparttable}
			\centering 		
			\begin{tabular}{|c|c|c|c|c|c|c|c|c|}			
				\hline                       
				\footnotesize
				\textbf{Receiver} & \footnotesize\textbf{Voltage} & \footnotesize\textbf{Impedance} & \footnotesize\textbf{Complexity}  & \footnotesize\textbf{Direct}  & \footnotesize\textbf{Frequency} & \footnotesize\textbf{Footprint} & \footnotesize\textbf{CMOS} & \footnotesize\textbf{Rate} \\ [0.5ex]
				& \footnotesize\textbf{bias} & \footnotesize\textbf{matching} & & \footnotesize\textbf{detection} & \footnotesize\footnotesize\textbf{(THz)} & \footnotesize\textbf{(mm\textsuperscript{2})} & \footnotesize\textbf{compatibility} & \footnotesize\textbf{(Gbits$^{-1}$)} \\
				\hline			                  	
				\footnotesize~GaAs \cite{tohme_2014}   & \footnotesize0.4 V & \footnotesize Few k$\Omega$ & \footnotesize Low & \footnotesize Yes & \footnotesize0.2 – 0.3  & \footnotesize Bulk & \footnotesize No & \footnotesize8.2 \\				
				\hline
				\footnotesize~MIMO \cite{simsek_140_2018} & \footnotesize1 V (4 channels) & \footnotesize- & \footnotesize High & \footnotesize No & \footnotesize0.140 & \footnotesize2.97 & \footnotesize Yes & \footnotesize0.8 \\
				\hline
				\footnotesize~SBD \cite{harter_2020} & \footnotesize zero-bias & \footnotesize- & \footnotesize High & \footnotesize Yes & \footnotesize0.2-0.3 & \footnotesize Bulk & \footnotesize No & \footnotesize115 \\
				\hline
				\footnotesize~Plasmonic MZM \cite{horst_2022} & \footnotesize1.34 V$^a$ & \footnotesize No & \footnotesize Medium & \footnotesize No & \footnotesize0.230 & \footnotesize Bulk$^*$ & \footnotesize Yes & \footnotesize$\sim$240 \\
				\hline
				\footnotesize~Plasmonic modulator \cite{ibili_2024} & \footnotesize zero-bias$^a$ & \footnotesize- & \footnotesize Medium & \footnotesize No & \footnotesize0.235 & \footnotesize Bulk$^*$ & \footnotesize Yes & \footnotesize80 \\
				\hline
				\footnotesize~This work & \footnotesize zero-bias & \footnotesize$\sim$100 $\Omega$ & \footnotesize Low & \footnotesize Yes & \footnotesize0.19–0.26 & \footnotesize0.018 & \footnotesize Yes & \footnotesize3 \\ [1ex]		    
				\hline								
			\end{tabular}
			\begin{tablenotes}
				\footnotesize
				\item $^*$ The Mach-Zehnder modulator (MZM) and plasmonic modulator have a footprint of 3$\times$10$^{-6}$ and 0.018 mm$^2$, respectively, but they require a laser (LO) and optical photodetector to complete the communication link. $^a$ The bias reported is applied to the modulators without considering the optical receiver's bias.			
			\end{tablenotes}
			\caption{Quantitative comparison of photodetectors operating at sub-THz frequencies.}				
		\end{threeparttable}
	\end{table*}
\end{savenotes}

In Table 1, we compare the performance of our developed receivers with those based on different technologies operating in the sub-THz range. We observe that our sub-THz receiver simultaneously achieves the main figures of merit, in contrast to other technologies. The direct detection mechanism allows the graphene receiver to have low complexity and minimal footprint for optimal performance. In addition, because of the integrated dipole antenna, we avoid using a bulky external horn antenna, and an additional optical amplifier. The zero-bias operation leads to low power consumption, low noise, and compatibility with CMOS technology. All the above-mentioned advantages provide a cost-effective solution and meets the size, weight, and power consumption (SWaP) requirements that 6G technologies must fulfill.
\\

An additional comparison with other graphene photodetectors in sub-THz range\cite{vicarelli_2012, bandurin_2018, bandurin_nc_2018, guo_2018, efetov_2018, blaikie_2019, lee_2020, jiang_2023, hong_2023, delgado-notario_2024, ryzhii_2024, ryzhii_2024_arxiv, santana_2022, rogalski_2022, spirito_2014, zak_2014, generalov_2017, auton_2017, liu_2018, liu_c_2018, delgado-notario_2022, ludwig_2024, joint_2024, bandurin_2024, roshan_2024} is presented in the Supplementary Table 3. Notably, none of them have demonstrated data stream detection. We stress that a higher data rate reception is potentially achievable by reporting a setup-limited bandwidth of 40 GHz for Device 7, for which it was not possible to measure the data stream detection owing to its low photosignal and responsivity. As previously mentioned, a bandwidth of 420 GHz was recently demonstrated at telecom frequencies for a PTE-based graphene detector.\cite{koepfli_2024,Koepfli_2023} Therefore, an intrinsically high bandwidth of graphene could be potentially achievable at sub-THz frequencies. 
\\

Our work demonstrates the potential of graphene receivers in the sub-THz range for wireless communications and aims for future investigations with other modulation schemes (e.g., four-level pulse-amplitude modulation) to achieve even higher data rates \cite{simsek_140_2018, giridhar_2024}. We anticipate that by engineering the sub-THz cavity by reducing its size using polaritonic resonators \cite{alfaro-mozaz_2017,hanan_2021,norenberg_2022,hanan_2022,obst_2023,lazzari_2020, xu_2023,castilla_2024,hanan_2024, tresguerres-mata_2024}, we can aim for higher optical bandwidths while maintaining high responsivity values. All of the above-mentioned factors enable graphene receivers to play a key role in future 6G wireless communications.
\\

\section*{Methods}

\subsection*{Fabrication of the graphene receivers}
\subsubsection*{Single Crystalline (SC) Graphene Growth and transfer}
\footnotesize
The growth process was carried out on a 35-$\mu$m thick copper (Cu) substrate using an Aixtron Black Magic cold-wall chemical vapor deposition (CVD) system. Prior to growth, the copper foil was electropolished to minimize surface roughness. The growth process began with an annealing step, during which the temperature was increased to 1000°C in an argon atmosphere for 30 min. Subsequently, the precursors, 50 sccm of H$_2$ and 30 sccm of CH$_4$ (0.01$\%$ CH$_4$ in Ar), were introduced to initiate nucleation. Growth was maintained for approximately 6 h at a constant pressure of 25 mbar. Subsequently, the CVD graphene was removed from the furnace and spin-coated with an A4-950K poly(methyl methacrylate) (PMMA) polymer from Microchemicals GmbH, which was subsequently cured at room temperature. The PMMA/SC-graphene/Cu composite was then transferred to 3 wt. $\%$ ammonium persulfate solution to etch away the copper foil. The PMMA/SC-graphene was rinsed by transferring it to deionized water and subsequently transferred onto a high-resistivity Si substrate. After draying at room temperature for one day, acetone/IPA baths were used to remove the PMMA layer from the top. 
\\

\subsubsection*{hBN encapsulated exfoliated graphene receivers}
\footnotesize
Graphene and hBN flakes are obtained through mechanical exfoliation from bulk highly oriented pyrolytic graphite (HOPG) and high-pressure, high-temperature-grown hBN single crystals, respectively, onto a 285 SiO$_2$/Si++ substrate. Following the identification of the appropriate thickness using atomic force microscopy (AFM), the top hBN (T-hBN) flake was picked using polycarbonate (PC) (5 wt. $\%$ in chloroform)/PDMS/glass stamp at 100°C. Subsequently, graphene and bottom hBN (B-hBN) were sequentially picked. The hBN-Gr-hBN stack/PC was then dropped at 180°C onto an SiO$_2$/ highly resistive Si double side polished substrate with prepatterned Cr/Au markers. Subsequently, PC was dissolved in chloroform for 30 min. One-dimensional source-drain electrodes were patterned using Elphy F50 Electron Beam Lithography (EBL) and Oxford Reactive Ion Etching (RIE) with SF$_6$ and O$_2$ gases. Subsequently, the metal deposition of Cr/Au was performed. The thickness of the metal electrode precisely matched that of the top hBN layer, and an additional dry etching step was introduced to isolate the photoactive channel. These procedures were conducted using 950 K-PMMA (from Microchemicals GmbH resist) as a lithographic mask. Subsequently, the sample was immersed overnight in acetone to facilitate a thorough clean lift-off process. Following this, another hBN flake (D-hBN) was dropped on top of the device, functioning as a dielectric layer that isolates the 5/200 nm thick (Cr/Au) dipole antenna. The dipole antenna was fabricated using Electron Beam Lithography (EBL) with a gap of ~400-500 nm.
The fabrication procedure for the TMD-Gr-hBN receiver described in this study remains consistent, with the exception of substituting the bottom hBN (B-hBN) layer with mechanically exfoliated WSe$_2$. WSe$_2$ flakes were exfoliated from the bulk and grown via chemical vapor transport (CVT).
\\

\subsubsection*{CVD Graphene receivers}
\footnotesize
Electron Beam Lithography (EBL) was used to pattern one-dimensional source and drain contacts, employing a 950 K PMMA mask on top of the SC CVD graphene on a highly resistive Si double side polished substrate. Subsequently, Reactive Ion Etching (RIE) with O$_2$/Ar was employed to etch graphene. We implement an H-shaped graphene channel to improve the responsivity,\cite{castilla_2019} and to reduce the contact resistance which is significant for CVD graphene-based receivers. A deposition of 5/20 nm of Cr/Au was performed, followed by overnight immersion in acetone to ensure an effective lift-off. Graphene undergoes a secondary patterning process using Electron Beam Lithography (EBL), followed by Reactive Ion Etching (RIE) to define the channel of the photodetector. A 40 nm-thick layer of alumina (Al$_2$O$_3$), which served as the dielectric and protective coating, was deposited at 250°C via Atomic Layer Deposition (ALD) (Cambridge Nanotech). Subsequently, an EBL patterning process was used to pattern a dipole antenna with a $\sim$400-500 nm gap, followed by the deposition of a 5/200 nm thick Cr/Au layer. The same procedure was employed for fabricating the CVD-graphene/hBN receivers, with ALD-grown alumina substituted with exfoliated D-hBN. D-hBN was picked and dropped using a PC/PDMS/glass stamp at temperatures of 100 and 180°C. 
\\

\subsection*{Electrical measurements}
\footnotesize
Electrical measurements were conducted at room temperature using a probe station from MPI Corporation. We use the probe station to check the devices at each fabrication step to verify the device integrity. The two-terminal resistance was measured by applying a 5 mV bias using a DAC, while the gate voltages for tuning the split gates were supplied by a Keithley 2640-B model.
\\

\subsection*{Optical simulations}
\footnotesize
Optical calculations were performed using the full-vector 3D finite-difference time-domain (FDTD) method with Lumerical software. The computational cell dimensions are 8×8×12.6 mm$^2$ with perfectly matched layer (PML) conditions applied to all boundaries. In the region occupied by the device, we employ a refined grid with 40 nm and 5 \textmu m in the lateral $x$ and $y$ dimensions, respectively, and 10 nm in the vertical, $z$, direction. The sub-THz source is modeled as an incident plane wave reaching the device through an open aperture of 7.5×7.5 \textmu m$^2$ with a spectral range of 40-400 GHz. The dimensions of the device layers are described in the SI and Fabrication section in Methods.
\\

\subsection*{Responsivity and NEP calculation}
\footnotesize
The external responsivity is given by: \\Responsivity = ($I$$_{\rm PTE}/P_{\rm in}$)$\times$($A$$_{\rm focus}$/$A$$_{\rm diff}$)\cite{vicarelli_2012, castilla_2019, castilla_2020, viti_2021}, where $P$$_{\rm in}$ is the power measured by the commercial Erickson PM5B power meter from Virginia Diodes, Inc. The measured $P$$_{\rm in}$ is approximately -1.65 dB (0.68 mW) at 249 GHz, with the graphene receiver (or power meter) placed 15 cm from the horn antenna output. $A_{\rm focus}$ is the experimental beam area at the measured wavelength and $A$$_{\rm diff}$ is the diffraction-limited spot size. We measure the photocurrent $I$$_{\rm PTE}$ from the output signal of the lock-in amplifier $V_{\rm LIA}$ considering $I$$_{\rm PTE} = \frac{2 \mathrm{\pi} \sqrt{2}}{4\xi} V_{\rm LIA}$\cite{vicarelli_2012, castilla_2019, castilla_2020, viti_2021}, where $\xi$ is the gain factor in V/A (given by the lock-in amplifier). We use the ratio $A$$_{\rm diff}/$$A$$_{\rm focus}$ for estimating the power reaching our photodetector since $A_{\rm diff}$ is the most reasonable value one can attain when considering the detector together with an optimized focusing system (e.g. using hemispherical lens) and it is widely used in the literature for comparing the performances among photodetectors \cite{vicarelli_2012, castilla_2019, castilla_2020, viti_2021}. This ratio is given by $A$$_{\rm diff}/$$A$$_{\rm focus} = \frac{w^2_{\rm 0,diff}}{w_{\rm 0,x} w_{\rm 0,y}}$. In order to obtain $w_{\rm 0,x}$ and $w_{\rm 0,y}$ we use our experimental observation that the photocurrent is linear in laser power and measure the photocurrent while scanning the device in the $x-$ and $y-$direction. Consequently, the photocurrent is described by Gaussian distributions $\propto \mathrm{e}^{-2x^2/w^2_{\rm 0,x}}$ and $\propto \mathrm{e}^{-2y^2/w^2_{\rm 0,y}}$, where $w_{\rm 0,x}$ and $w_{\rm 0,y}$ are the respectively obtained spot sizes (related to the standard deviation via $\sigma = w_0/2$ and to the FWHM = $\sqrt{2 \ln(2)}w_0$) \cite{vicarelli_2012, castilla_2019, castilla_2020, viti_2021}. We usually achieve $w_{\rm 0,x}$ = 10.1 mm and $w_{\rm 0,y}$ = 8.8 mm at $\lambda$ = 1407.5 mm (0.213 THz). We note that for the measurements performed to determine the responsivity, we did not employ focusing elements, such as lenses. Therefore, the sub-THz beam emitted from the horn antenna toward the graphene receiver, placed at a distance of 15 cm, is highly divergent. For the diffraction-limited spot, we consider $w_{\rm 0,diff} = \frac{\lambda}{\mathrm{\pi}}$, with $\lambda$ the sub-THz wavelength. The diffraction-limited area is hence taken as $A$$_{\rm diff} = \mathrm{\pi} w_{\rm 0, diff}^2 = \lambda^2/\mathrm{\pi}$. Additionally, the noise-equivalent power (NEP) that characterizes the sensitivity of the photodetector is defined as NEP $ = I_{\rm noise}/$Responsivity and considering that our unbiased photodetector has a very low noise current that is limited by Johnson noise, we use a noise spectral density $I_{\rm noise} = \sqrt{\frac{4k_{\rm B}T}{R_{\rm D}}}$, where $k_{\rm B}$ corresponds to the Boltzmann constant, $T$ is the operation temperature (300 K) and $R_{\rm D}$ the device resistance. 
\section*{Acknowledgments}
\small
F.H.L.K. acknowledges financial support from the Spanish Ministry of Economy and Competitiveness, through the “Severo Ochoa” Programme for Centres of Excellence in R$\&$D (SEV-2015-0522), support by Fundacio Cellex Barcelona, Generalitat de Catalunya through the CERCA program,  and  the Agency for Management of University and Research Grants (AGAUR) 2017 SGR 1656.  Furthermore, the research leading to these results has received funding from the European Union Seventh Framework Programme under grant agreement no.785219 and no. 881603 Graphene Flagship for Core2 and Core3. K.J.T. acknowledges funding from the European Union’s Horizon 2020 research and innovation program under Grant Agreement No. 101125457 (ERC CoG "EQUATE"). ICN2 is funded by the CERCA Programme / Generalitat de Catalunya and supported by the Severo Ochoa Centres of Excellence programme, Grant CEX2021-001214-S, funded by MCIN / AEI / 10.13039.501100011033. Results incorporated in this standard received funding from the European Union’s Horizon 2020 research and innovation programme under grant agreement No. 101034929. S.C., S.M., K.P.S. and F.H.L.K. acknowledge PDC2022-133844-I00, funded by MCIN/AEI/10.13039/501100011033 and by the “European Union NextGenerationEU/PRTR". S.C., K.P.S. and F.H.L.K acknowledge funding by the European Union (ERC,TERACOMM, 101113529). Views and opinions expressed are however those of the author(s) only and do not necessarily reflect those of the European Union or the European Research Council Executive Agency. Neither the European Union nor the granting authority can be held responsible for them.  

\section*{Author contributions}
\small
S.C., S.M. and F.H.L.K. conceived the project. K.P.S. and S.C. fabricated the devices. S.K. fabricated device 7. S.C., S.M., K.P.S., S.K. and L.K. performed the experiments. I.V. and E.L. performed the simulations. S.C. assisted in the modelling. S.M. performed preliminary optical simulations. S.M. designed the evaluation kit. S.C., K.P.S., S.K., L.K., K.-J.T. and F.H.L.K. wrote the manuscript. R.d.B., E.R. and K.-J.T. assisted with measurements and discussion of the results. K.W. and T.T. synthesized the hBN crystals. S.T. synthesized the TMD crystals. E.L., K.-J.T., J.L. and F.H.L.K. supervised the work and discussed the results. All authors contributed to the scientific discussion and manuscript revisions. K.P.S. and S.C. contributed equally to the work. 

\section*{Data availability}
{\small
The data that support the plots within this paper and other findings of this study are available from the corresponding author upon reasonable request.\\ 

}

\section*{Competing interests}
{\small
The authors declare no competing interests.
}

\clearpage

\clearpage

\newpage


\end{document}